\documentclass{article}
\usepackage{amsmath}
\usepackage{graphics}

\input{tcilatex}
\begin{document}

\author{Chih-Yuan Tseng\thanks{%
E-mail address: ct7663@csc.albany.edu} and Ariel Caticha\thanks{%
E-mail address: Ariel@albany.edu} \\
Department of physics, University at Albany-SUNY\\
Albany, NY 12222 USA}
\title{Maximum Entropy Approach to the Theory of Simple Fluids\thanks{%
Presented at MaxEnt 2003, the 23rd International Workshop on Bayesian
Inference and Maximum Entropy Methods (August 3-8, 2003, Jackson Hole, WY,
USA)} }
\date{}
\maketitle

\begin{abstract}
We explore the use of the method of Maximum Entropy (ME) as a technique to
generate approximations. In a first use of the ME method the
\textquotedblleft exact\textquotedblright\ canonical probability
distribution of a fluid is approximated by that of a fluid of hard spheres;
ME is used to select an optimal value of the hard-sphere diameter. These
results coincide with the results obtained using the Bogoliuvob variational
method. A second more complete use of the ME\ method leads to a better
descritption of the soft-core nature of the interatomic potential in terms
of a statistical mixture of distributions corresponding to hard spheres of
different diameters. As an example, the radial distribution function for a
Lennard-Jones fluid (Argon) is compared with results from molecular dynamics
simulations. There is a considerable improvement over the results obtained
from the Bogoliuvob principle.
\end{abstract}

\section{Introduction}

The method of Maximum Entropy (ME) is designed to solve the general problem
of updating from a prior probability distribution (which often happens to be
a uniform distribution) to a posterior distribution when new information in
the form of constraints becomes available. Of all the distributions
satisfying the constraints the preferred posterior is that which is closest
to the prior in the sense that it represents the least change of beliefs 
\cite{Caticha03}.

This suggests that ME can be used to tackle a different kind of problem.
Even if a distribution is known and it accurately reflects our beliefs it
may still turn out to be too complicated to be useful in practice. We may
need to find a more tractable approximation. The idea is to identify a
family of tractable trial distributions and then select that member which is
closest to the \textquotedblleft exact\textquotedblright\ distribution by
maximizing the appropriate relative entropy. Notice that this is not a
problem that can be tackled using the more restricted version of the maximum
entropy method usually known as MaxEnt. The reason is that MaxEnt only
allows updating from an underlying physical measure and not from a general
prior distribution.

The purpose of this paper is to develop the ME method as a technique to
generate approximations. In previous work \cite{Tseng02} we have shown that
a number of well-known variational approximation schemes (the Bogoliubov
variational method, the Kohn-Sham-Hohenberg density functional formalism,
and even the variational method of quantum mechanics of which the
Hartree-Fock method is an important special case) can be derived as special
cases of the ME method. In \cite{Tseng02} the ME method was used to develop
a mean-field approximation for classical fluids in which the tractable
family of distributions is obtained by replacing the interactions between
molecules by an effective external mean field. The effects of the long-range
attractions are described well but the short-range repulsions are badly
misrepresented.

In section 2 the treatment of short-range repulsions is considerably
improved by selecting as our tractable model a fluid of hard spheres \cite%
{BarkerHenderson76}-\cite{Kalikmanov02}. The ME method is then used (section
3) to select the optimal value of the hard-sphere diameter. This is
equivalent to applying the Bogoliuvob variational principle and reproduces
the results obtained by Mansoori et al. \cite{Mansoori69}.

An alternative approach to the study of fluids is through thermodynamic
perturbation theory which yields results formally similar to the variational
method. In such schemes the actual interatomic potential $u$ is replaced by $%
u_{0}+\delta u$, where $u_{0}$ represents the strong short-range repulsion
and $\delta u$ is a long-range attraction treated as a perturbation.
Remarkably, the structure of the fluid for liquid densities is dominated by
the repulsive interactions. The effects of the attraction $\delta u$ are
averaged over many molecules and do not appreciably affect the correlations
among molecules. As a result, the first-order perturbation theory is quite
accurate. At lower densities, however, higher-order corrections must be
included. Since $u_{0}$ is not itself a tractable potential the usual
approach is to replace it by a hard-sphere potential $u_{hs}$. Several
suggestions of how to separate $u$ into $u_{0}$ and $\delta u$, of where and
how to replace $u_{0}$ by $u_{hs}$, and how to choose the best hard-sphere
diameter have been proposed. The most successful are the theory of Barker
and Henderson \cite{BarkerHenderson76} and, particularly, the theory of
Weeks, Chandler and Andersen (WCA) \cite{WCA} which succeeds in using the
hard-sphere $u_{hs}$ while effectively representing the effects of the
soft-core potential $u_{0}$. For a recent discussion of some of the
strengths and limitations of the perturbative approach see \cite{Germain02}.

A clear advantage of the variational and the ME methods is that the
important and yet somewhat ad hoc nature of the separation of $u$ into $%
u_{0} $ and $\delta u$, and of the choice of a hard-sphere diameter are
eliminated. On the other hand, the variational approach fails to take the
softness of the repulsive core into account, and this leads, in the end, to
results that are inferior to the perturbative approaches particularly at
high temperatures.

The traditional variational approach allows one to select a single optimal
diameter; all non-optimal values are ruled out. But, as discussed in \cite%
{Caticha03, Caticha00}, the ME method allows one to proceed further and
quantify the extent to which non-optimal values should contribute. In this
more complete use of the ME method, presented in section 4, the
\textquotedblleft exact\textquotedblright\ probability distribution of the
fluid is approximated not by that of a gas of hard spheres with the optimal
diameter but by a statistical mixture of distributions corresponding to hard
spheres of different diameters. This is a rather simple and elegant way to
take proper account of the fact that the actual atoms are not hard spheres
but have a soft core. The full ME analysis leads to significant improvements
over the variational method.

When faced with the difficulty of dealing a system described by an
intractable Hamiltonian, the traditional approach has been to consider a
similar albeit idealized system described by a simpler more tractable
Hamiltonian. The approach we have followed here departs from this tradition:
our goal is not to identify an approximately similar system but rather to
identify an approximately similar probability distribution. The end result
of the ME approach is a probability distribution which is a sum over
distributions corresponding to different hard-sphere diameters. While each
term in the sum is of a form that can be associated to a real hard-sphere
gas, the sum itself is not of the form $\exp -\beta H$, and cannot be
interpreted as describing any physical system.

In section 5 we test our method by comparing its predictions for Argon gas
with the numerical molecular dynamics simulation data obtained by Verlet 
\cite{Verlet68}. We find that the ME predictions for thermodynamic variables
and for the radial distribution function are considerable improvements over
the Bogoliuvob variational result, and are comparable to the perturbative
results. Finally, our conclusions and some remarks on further improvements
are presented in section 6.

\section{Approximation by hard spheres}

We consider a simple fluid composed of $N$ single atom molecules described
by the Hamiltonian \ 

\begin{equation}
H(q_{N})=\sum_{i=1}^{N}\,\frac{p_{i}^{2}}{2m}+U\quad \text{with}\quad
U=\sum_{i>j}^{N}u(r_{ij})\,,  \label{Actual-H}
\end{equation}%
where $q_{N}=\{p_{i},r_{i};\;i=1,...,N\}$ and the many-body interactions are
approximated by a pair interaction, $u(r_{ij})$ where $r_{ij}=\left\vert
r_{i}-r_{j}\right\vert $. The probability that the positions and momenta of
the molecules lie within the phase space volume $dq_{N}$ is given by
canonical distribution, and 
\begin{equation}
P(q_{N})\,dq_{N}=\frac{1}{Z}e^{-\beta H(q_{N})\,}\,dq_{N}~,
\label{exact dist}
\end{equation}%
where

\begin{equation}
dq_{N}=\frac{1}{N!\,h^{3N}}\prod_{i=1}^{N}d^{3}p_{i}d^{3}r_{i}\quad \text{and%
}\quad Z=\int dq_{N}\text{ }e^{-\beta H(q_{N})}\,.
\end{equation}

The difficulty, of course, is that this distribution is easy to write down
but very difficult to use. We must replace $P$ by a more tractable
distribution. To proceed one must recognize the two features of the
interaction potential $u$ that seem to be relevant for explaining a wide
variety of fluid properties; they are the strong short-distance repulsion
and the weaker long-distance repulsion.

To account for the short-distance repulsion we consider the probability
distribution corresponding to a gas of $N$ hard spheres of diameter $r_{d}$.
The Hamiltonian is

\begin{equation}
H_{hs}(q_{N}\left\vert r_{d}\right. )=\sum_{i=1}^{N}\,\frac{p_{i}^{2}}{2m}%
+U_{hs}\quad \text{with}\quad U_{hs}=\sum_{i>j}^{N}u_{hs}(r_{ij}|r_{d})~,
\label{Hhs}
\end{equation}%
where 
\begin{equation}
u_{hs}(r\left\vert r_{d}\right. )=\left\{ 
\begin{array}{ccc}
0 & \text{for} & r\geq r_{d} \\ 
\infty & \text{for} & r<r_{d}%
\end{array}%
\right.
\end{equation}%
The corresponding probability distribution is 
\begin{equation}
P_{hs}(q_{N}\left\vert r_{d}\right. )=\frac{1}{Z_{hs}}e^{-\beta
H_{hs}(q_{N}\left\vert r_{d}\right. )}\,.  \label{Phs}
\end{equation}%
The partition function and the free energy $F_{hs}(T,V,N\left\vert
r_{d}\right. )$ are

\begin{equation}
Z_{hs}=\int dq_{N}\,\text{\ }e^{-\beta H_{hs}(q_{N})}\,\overset{\limfunc{def}%
}{=}e^{-\beta F_{hs}(T,V,N\left\vert r_{d}\right. )}\,.
\end{equation}

Two objections that can be raised to the choice of $P_{hs}$ are, first, that
it does not take the long-range interactions into account; this is a point
to which we will return later. Second, and this is a more serious problem,
it is not clear that $P_{hs}$ is a tractable distribution at all. Indeed,
the exact solution to the problem of $N$ hard spheres is not known. However,
there exist analytical approximations that are in reasonably good agreement
with numerical simulations. We will therefore assume that for all practical
purposes $P_{hs}$ is a tractable distribution.

For hard spheres the radial correlation function can be calculated within
the approximation of Percus and Yevick \cite{PercusYevick58, Wertheim63} or,
alternatively, from the scaled-particle theory \cite{Reiss59}. The equation
of state can then be computed in two alternative ways, either from the
so-called \textquotedblleft pressure\textquotedblright\ equation, or from
the \textquotedblleft compressibility\textquotedblright\ equation but the
two results do not agree. It has been found that better agreement with
simulations and with virial coefficients is obtained taking an average of
the two results with weights 1/3 and 2/3 respectively. The result is the
Carnahan and Starling equation of state, \cite{BarkerHenderson76}-\cite%
{Kalikmanov02}

\begin{equation}
\left. \frac{PV}{Nk_{B}T}\right\vert _{hs}=\frac{1+\eta +\eta ^{2}-\eta ^{3}%
}{\left( 1-\eta \right) ^{3}}\text{ },  \label{HS-P}
\end{equation}%
where $\eta =\frac{1}{6}\pi \rho r_{d}^{3}$ with $\rho =N/V$. The free
energy, derived by integrating the equation of state, is

\begin{equation}
F_{hs}(T,V,N\left\vert r_{d}\right. )=Nk_{B}T\left[ -1+\ln \rho \Lambda
_{T}^{3}+\frac{4\eta -3\eta ^{2}}{\left( 1-\eta \right) ^{2}}\right] \text{ }
\label{HS-Free energy}
\end{equation}%
where $\Lambda _{T}=(2\pi \hbar ^{2}/mk_{B}T)^{1/2}$, and the entropy is

\begin{equation}
S_{hs}\left[ P_{hs}\right] =\left( \frac{\partial F_{hs}}{\partial T}\right)
_{N,V}=\frac{F_{hs}}{T}+\frac{3}{2}Nk_{B}\text{ }.  \label{HS-S}
\end{equation}%
It must be remembered that these expressions are not exact. They are quite
reasonable approximations for all densities up to almost crystalline
densities (about $\eta \approx 0.5$). However, they fail to predict the
face-centered-cubic phase when $\eta $ is in the range from $0.5$ up the
close-packing value of $\eta _{cp}\approx 0.74$.

\section{ME determination of the best hard-sphere diameter}

Next we address the question of which is the best $P_{hs}$. Which is the
best hard-sphere diameter? According to the ME method \cite{Caticha03}, the
preferred trial $P_{hs}(q_{N}|r_{d})$ is that which is closest to the
\textquotedblleft exact\textquotedblright\ $P(q_{N})$. It is found by
maximizing the relative entropy,

\begin{equation}
\mathcal{S}\left[ P_{hs}|P\right] =-\int dq_{N}\,P_{hs}(q_{N}|r_{d})\log 
\frac{P_{hs}(q_{N}|r_{d})}{P(q_{N})}\text{ }\leq 0\text{.}
\end{equation}%
Substituting eqs.(\ref{exact dist}) and (\ref{Phs}) we obtain

\begin{equation}
\mathcal{S}\left[ P_{hs}|P\right] =\beta \left[ F-F_{hs}-\langle
U-U_{hs}\rangle _{hs}\right] \leq 0\,,  \label{S1}
\end{equation}%
where $\langle \cdots \rangle _{hs}$ is the average computed over the
hard-sphere distribution $P_{hs}(q_{N}|r_{d})$. Next rewrite eq.(\ref{S1}) as

\begin{equation}
F\leq F_{U}\overset{\limfunc{def}}{=}F_{hs}+\langle U-U_{hs}\rangle _{hs}%
\text{ },
\end{equation}%
and use 
\begin{equation}
\langle U-U_{hs}\rangle _{hs}=\frac{1}{2}\int d^{3}rd^{3}r^{\prime }\left[
\,u(r-r^{\prime })-u_{hs}(r-r^{\prime }|r_{d})\right] n_{hs}^{\left(
2\right) }(r,r^{\prime })~,
\end{equation}%
\ where $n_{hs}^{\left( 2\right) }(r,r^{\prime })=\langle n^{\left( 2\right)
}(r,r^{\prime })\rangle _{hs}$ and 
\begin{equation}
\hat{n}^{(2)}(r,r^{\prime })\overset{\limfunc{def}}{=}\sum_{i\neq j}\,\delta
(r-r_{i})\,\delta (r^{\prime }-r_{j})=\hat{n}(r)\hat{n}(r^{\prime })-\hat{n}%
(r)\delta (r-r^{\prime })\;,
\end{equation}%
is the two-particle density distribution. But $u_{hs}(r-r^{\prime }|r_{d})=0$
for $\left\vert r-r^{\prime }\right\vert \geq r_{d}$ while $n_{hs}^{\left(
2\right) }(r,r^{\prime })=0$ for $\left\vert r-r^{\prime }\right\vert \leq
r_{d}$, therefore

\begin{equation}
F_{U}=F_{hs}+\langle U\rangle _{hs}\quad \text{with}\quad \langle U\rangle
_{hs}=\frac{1}{2}N\rho \int d^{3}r\,u(r)g_{hs}(r\left\vert r_{d}\right. )\,,
\label{F_U}
\end{equation}%
where we have assumed that the fluid is homogeneous, $n_{hs}^{\left(
2\right) }(r,r^{\prime })=n_{hs}^{\left( 2\right) }(\left\vert r-r^{\prime
}\right\vert )$, and introduced the radial distribution function $g_{hs}$
defined by 
\begin{equation}
n_{hs}^{\left( 2\right) }(r)\overset{\func{def}}{=}\rho
^{2}g_{hs}(r\left\vert r_{d}\right. ).
\end{equation}

Maximizing $\mathcal{S}\left[ P_{hs}|P\right] $ is equivalent to minimizing $%
F_{U}$ over all diameters $r_{d}$. Thus, the variational approximation to
the free energy is

\begin{equation}
F\left( T,V,N\right) \approx F_{U}(T,V,N\left\vert r_{m}\right. )\overset{%
\limfunc{def}}{=}\min_{r_{d}}F_{U}(T,V,N\left\vert r_{d}\right. )\,,
\label{mini-Fu}
\end{equation}%
where $r_{m}$ is the optimal diameter. Notice that the approximation does
not consist of merely replacing the free energy $F$ by a hard-sphere free
energy $F_{hs}$ which does not include the effects of long range attraction; 
$F$ is approximated by $F_{U}(r_{m})$ which includes attraction effects
through the second term in eq.(\ref{F_U}). This addresses the first of the
two objections mentioned earlier. Indeed, the real fluid with interactions
given by $u$ is not being replaced by a hard-sphere fluid with interactions
given by $u_{hs}$; it is just the probability distribution that is being
replaced in this way. The internal energy is approximated by $\langle
H\rangle _{hs}=\frac{3}{2}Nk_{B}T+\langle U\rangle _{hs}$ and not by $%
\langle H_{hs}\rangle _{hs}=\frac{3}{2}Nk_{B}T$.

To complete the solution one needs the radial distribution function $%
g_{hs}(r\left\vert r_{d}\right. )$. A numerical solution by Throop and
Bearman is tabulated in \cite{Throop65}. Alternatively it can be written in
terms of the Laplace transform of $rg_{hs}(r\left\vert r_{d}\right. )$ \cite%
{Wertheim63},

\begin{equation}
G(s)=\int_{0}^{\infty }dx~xg_{hs}(xr_{d}|r_{d})e^{-sx}=\frac{sL(s)}{12\eta %
\left[ L(s)+S(s)e^{s}\right] }
\end{equation}%
where $x$ is a dimensionless variable $x=r/r_{d}$, 
\begin{equation}
L(s)=12\eta \left[ \left( 1+\frac{1}{2}\eta \right) s+\left( 1+2\eta \right) %
\right] ~,
\end{equation}%
and

\begin{equation}
S(s)=\left( 1-\eta \right) ^{2}s^{3}+6\eta \left( 1-\eta \right)
s^{2}+18\eta ^{2}s-12\eta \left( 1+2\eta \right) \text{ }.
\end{equation}%
To make use of this solution it is convenient to write $\langle U\rangle
_{hs}$ in terms of $V(s)$, the inverse Laplace transform of $ru(r)$, 
\begin{equation}
xu(xr_{d})=\int_{0}^{\infty }ds~V(s)e^{-sx}~,
\end{equation}%
which gives

\begin{equation}
\langle U\rangle _{hs}=12N\eta \int_{0}^{\infty }ds\text{~}V(s)G(s)\,,
\label{F_U-LT}
\end{equation}%
For a Lennard-Jones potential, 
\begin{equation}
u(r)=4\varepsilon \left[ \left( \frac{\sigma }{r}\right) ^{12}-\left( \frac{%
\sigma }{r}\right) ^{6}\right]  \label{Lennard-Jones}
\end{equation}%
we have

\begin{equation}
V(s)=4\varepsilon \left[ \left( \frac{\sigma }{r_{d}}\right) ^{12}\frac{%
s^{10}}{10!}-\left( \frac{\sigma }{r_{d}}\right) ^{6}\frac{s^{4}}{4!}\right]
~.
\end{equation}%
Finally, it remains to minimize $F_{U}$ in eq.(\ref{F_U}). This is done
numerically; an explicit example for Argon is calculated in section 5.\ 

As mentioned in the Introduction a problem with the approach outlined above
is that it fails to take the softness of the repulsive core into account.
This flaw is manifested in a less satisfactory prediction of thermodynamic
variables at high temperatures, and also, as will be shown by the numerical
calculations in section 4, in a poor prediction of the radial distribution
function. However, as we shall see in the following sections, the ME method
does not stop at the mere selection of the best diameter. A more complete ME
analysis offers significant improvements over the variational method.

\section{A more complete ME analysis}

The ME method as pursued in the last section has led us to determine an
optimal value of the hard-sphere diameter. Next we ask to what extent do we
believe that the correct selection should be $r_{d}=r_{m}$ rather than $%
r_{d}=r_{m}+\delta r$. To what extent are values $\delta r\neq 0$ ruled out
by the ME method? \cite{Caticha03, Caticha00} This question is an inquiry
about the probability of $r_{d}$, $P_{d}(r_{d})$. Thus, we are uncertain not
just about $q_{N}$ given $r_{d}$, but also about the right $r_{d}$ and what
we actually seek is the joint probability of $q_{N}$ and $r_{d}$, $%
P_{J}(q_{N},r_{d})$. Once this joint distribution is obtained our best
assessment of the distribution of $q_{N}$ should be given by the marginal
over $r_{d}$, 
\begin{equation}
\bar{P}_{hs}(q_{N})\overset{\func{def}}{=}\int dr_{d}\text{ }%
P_{J}(q_{N},r_{d})=\int dr_{d}\text{~}P_{d}(r_{d})P_{hs}(q_{N}|r_{d})\text{ }
\label{Pbar(q)}
\end{equation}%
By recognizing that diameters other than $r_{m}$ are not ruled out and that
a more honest representation is an average over all hard-sphere diameters we
are effectively replacing the hard spheres by a soft-core potential.

Next, we make a second use of the ME method to obtain $P_{J}(q_{N},r_{d})$.
Of all distributions within the family $%
P_{J}(q_{N},r_{d})=P_{d}(r_{d})P_{hs}(q_{N}|r_{d})$ we seek that which is
closest to the prior $m(q_{N},r_{d})$ over the space $\{q_{N},r_{d}\}$. But
what is $m(q_{N},r_{d})$? We want the marginal distribution $\bar{P}%
_{hs}(q_{N})$, eq.(\ref{Pbar(q)}), as close as possible to the
\textquotedblleft exact\textquotedblright\ distribution $P(q_{N})$ in eq.(%
\ref{exact dist}). This is achieved if we set $m(q_{N},r_{d})=P(q_{N})\mu
(r_{d}).$ Furthermore, we can assign $\mu (r_{d})$ noting that $r_{d}$ is
not a physical quantity, its only meaning derives from being a parameter
appearing in the distribution $P_{hs}(q_{N}|r_{d}).$ Then there is a unique
natural distance $d\ell ^{2}=\gamma (r_{d})dr_{d}^{2}$ in the space of $%
r_{d} $s which is given by the Fisher-Rao metric,

\begin{equation}
\gamma (r_{d})=\int dq_{N}\,P_{hs}(q_{N}\left\vert r_{d}\right. )\left( 
\frac{\partial \log P_{hs}(q_{N}\left\vert r_{d}\right. )}{\partial r_{d}}%
\right) ^{2}.  \label{FR-metric}
\end{equation}%
Therefore $\mu (r_{d})=\gamma ^{1/2}(r_{d})$. Therefore the joint $%
P_{J}(q_{N},r_{d})$, or equivalently the diameter distribution $P_{d}(r_{d})$%
, is determined by maximizing the entropy

\begin{equation}
\sigma \left[ P_{d}\right] =-\int dq_{N}\,dr_{d}\text{ }%
P_{d}(r_{d})P_{hs}(q_{N}|r_{d})\log \frac{P_{d}(r_{d})P_{hs}(q_{N}|r_{d})}{%
\gamma ^{1/2}(r_{d})P(q_{N})}.\text{ }
\end{equation}%
For sake of simplicity, one can reorginize this entropy to

\begin{equation}
\sigma \left[ P_{d}\right] =-\int \,dr_{d}P_{d}(r_{d})\log \frac{P_{d}(r_{d})%
}{\gamma ^{1/2}(r_{d})}+\int \,dr_{d}P_{d}(r_{d})\mathcal{S}\left[ P_{hs}|P%
\right] \text{ },  \label{sigma Pd}
\end{equation}%
where $\mathcal{S}\left[ P_{hs}|P\right] $ is given eqs.(\ref{S1}) and (\ref%
{F_U}), 
\begin{equation}
\mathcal{S}\left[ P_{hs}|P\right] =\beta \left( F-F_{U}\right) =\beta \left(
F-F_{hs}-\langle U\rangle _{hs}\right) ~.
\end{equation}%
Maximizing $\sigma \left[ P_{d}\right] $ over variations $\delta P_{d}$
subject to a normalization constraint gives,

\begin{equation}
P_{d}(r_{d})dr_{d}=\frac{e^{\mathcal{S}\left[ P_{hs}|P\right] }}{\zeta }%
\gamma ^{1/2}\left( r_{d}\right) dr_{d}=\frac{e^{-\beta F_{U}}}{\zeta _{U}}%
\gamma ^{1/2}\left( r_{d}\right) dr_{d}\text{ },  \label{Pd(rd)}
\end{equation}%
where the partition functions $\zeta $ and $\zeta _{U}$ are given by

\begin{equation}
\zeta =e^{\beta F}\zeta _{U}\quad \text{with}\quad \zeta _{U}=\int dr_{d}%
\text{ }\gamma ^{1/2}\left( r_{d}\right) e^{-\beta F_{U}}\text{ }.
\end{equation}%
The remaining problem in the above equations is the calculation of the
Fisher-Rao measure $\gamma ^{1/2}$ and this is conveniently done by
considering the entropy of $P_{hs}(q_{N}\left\vert r_{d}^{\prime }\right. )$
relative to $P_{hs}(q_{N}\left\vert r_{d}\right. )$,

\begin{equation}
\mathcal{S}\left[ P_{hs}\left( q_{N}\left\vert r_{d}^{\prime }\right.
\right) \left\vert P_{hs}\left( q_{N}\left\vert r_{d}\right. \right) \right. %
\right] =-\int dq_{N}\,\text{ }P_{hs}(q_{N}\left\vert r_{d}^{\prime }\right.
)\log \frac{P_{hs}(q_{N}\left\vert r_{d}^{\prime }\right. )}{%
P_{hs}(q_{N}\left\vert r_{d}\right. )}~\text{. }  \label{S2}
\end{equation}%
A straightforward differentiation shows that

\begin{equation}
-\left. \frac{\partial ^{2}\mathcal{S}\left[ P_{hs}(q_{N}\left\vert
r_{d}^{\prime }\right. )\left\vert P_{hs}(q_{N}\left\vert r_{d}\right.
)\right. \right] }{\partial r_{d}^{\prime 2}}\right\vert _{r_{d}^{\prime
}=r_{d}}=\gamma (r_{d})\text{ }.  \label{det g}
\end{equation}%
Substituting the distributions $P_{hs}(q_{N}\left\vert r_{d}^{\prime
}\right. )$ and $P_{hs}(q_{N}\left\vert r_{d}\right. )$ into eq.(\ref{S2})
gives

\begin{equation}
\mathcal{S}\left[ P_{hs}(q_{N}|r_{d}^{\prime })|P_{hs}(q_{N}\left\vert
r_{d}\right. )\right] =\beta \left[ F_{hs}\left( r_{d}\right) -F_{hs}\left(
r_{d}^{\prime }\right) -\langle U_{hs}\left( r_{d}\right) \rangle
_{r_{d}^{\prime }}+\langle U_{hs}\left( r_{d}^{\prime }\right) \rangle
_{r_{d}^{\prime }}\right] \text{ },
\end{equation}%
where $\langle \cdots \rangle _{r_{d}^{\prime }}$ is the average over $%
P_{hs}(q_{N}|r_{d}^{\prime })$. As we argued above eq.(\ref{F_U}) the
expectation of the potential energy $\langle U_{hs}\left( r_{d}^{\prime
}\right) \rangle _{r_{d}^{\prime }}$ vanishes because the intermolecular
potential $u(r\left\vert r_{d}^{\prime }\right. )$ is zero in the region $%
r>r_{d}^{\prime }$ where the radial distribution function $%
g_{hs}(r\left\vert r_{d}^{\prime }\right. )$ is not zero. A similar argument
shows that $\langle U_{hs}\left( r_{d}\right) \rangle _{r_{d}^{\prime }}=0$
when $r_{d}^{\prime }\geq r_{d}$. However, when $r_{d}^{\prime }\leq r_{d}$
the expectation $\langle U_{hs}\left( r_{d}\right) \rangle _{r_{d}^{\prime
}} $ diverges, $\mathcal{S}$ is not defined and eq.(\ref{det g}) is not
applicable. We can argue our way out of this quandary by pointing out that
the divergence is a consequence of the unphysical idealization involved in
taking a hard-sphere potential seriously. For more realistic continuous
potentials the distance between $r_{d}^{\prime }=r_{d}+dr_{d}$ and $r_{d}$
is the same as the distance between $r_{d}^{\prime }=r_{d}-dr_{d}$ and $%
r_{d} $. We can then always choose $r_{d}^{\prime }\geq r_{d}$ and define $%
\gamma (r_{d})$ in eq.(\ref{det g}) as the limit $r_{d}^{\prime
}=r_{d}+0^{+} $. Then, using eq.(\ref{HS-Free energy}), we have

\begin{equation}
\gamma (r_{d})=\beta \left. \frac{\partial ^{2}F_{hs}\left( r_{d}^{\prime
}\right) }{\partial r_{d}^{\prime 2}}\right\vert _{r_{d}^{\prime
}=r_{d}+0^{+}}\text{ }=N\pi \rho r_{d}\frac{4+9\eta -4\eta ^{2}}{\left(
1-\eta \right) ^{4}}~.  \label{det g-exact}
\end{equation}

To summarize, the distribution of diameters $P_{d}(r_{d})$ is given by eq.(%
\ref{Pd(rd)}) with $F_{U}$ given by eqs. (\ref{F_U}, \ref{HS-Free energy}, %
\ref{F_U-LT}) and $\gamma $ given by (\ref{det g-exact}). Our best
approximation to the \textquotedblleft exact\textquotedblright\ $P(q_{N})$
is the $\bar{P}_{hs}(q_{N})$ given in eq.(\ref{Pbar(q)}). The corresponding
best approximation to the radial distribution function is

\begin{equation}
\bar{g}_{hs}(r)=\int dr_{d}\text{~}P_{d}(r_{d})g_{hs}(r\left\vert
r_{d}\right. )\text{ . }  \label{g(r)-approx1}
\end{equation}

When intermolecular interactions are given by two-body potentials the
equation of state, the free energy, the internal energy, and other
thermodynamic functions can be expressed in terms of the radial distribution
function. Therefore an improved $g(r)$ leads to improved estimates for all
other quantities.

However, there is a problem. For large $N$ the distribution $%
P_{d}(r_{d})\sim \exp -\beta F_{U}$ is very sharply peaked about the maximum
attained at the optimal diameter $r_{m}$ because $F_{U}$ is an extensive
quantity $F_{U}\propto N$. This result must be interpreted with care: when
choosing a single optimal diameter for a macroscopic fluid sample we find
that the ME confers overwhelming probability to the optimal value. This is
not surprising. The same thing happens when we calculate the global
temperature or density of a macroscopic sample: the ME method predicts that
fluctuations about the expected value are utterly negligible. And yet
fluctuations can be important. For example, for smaller fluid samples, or
when we consider the local behavior of the fluid, fluctuations are not
merely observable but can be large. Local fluctuations can be appreciable
while global fluctuations remain negligible.

The question then, is whether these local fluctuations are relevant to the
particular quantities we want to calculate. We argue that they are. The
radial distribution function $g(r)$ is the crucial quantity from which all
other thermodynamic variables are computed. But from its very definition -- $%
g(r)$ is the probability that given one atom at a given fixed point, another
atom will be found at a distance $r$ -- it is clear that $g(r)$ refers to
purely local behavior and should be affected by local fluctuations. To the
extent that the optimal $r_{m}$ depends on temperature and density we expect
that local temperature and/or density fluctuations would induce local
diameter fluctuations as well.

It is important to note that unlike density fluctuations, the local diameter
fluctuations are not real. They cannot be: the hard spheres do not exist
except in our minds. The diameter fluctuations are merely a representation
of our uncertainty about which $r_{d}$ to choose.

The extended analysis in this section does not yet allow us to pursue the
question of local fluctuations in a satisfactory manner. For the purpose of
this paper, however, we can quickly estimate the effects of local
fluctuations by assuming that the effective number of atoms $N_{\func{eff}}$
that are locally relevant to the calculation of $g(r)$ is smaller than $N$.
We deal with an effectively smaller fluid sample. The actual calculation of $%
N_{\func{eff}}$ will be pursued elsewhere; it is not difficult to see that
the ME method itself still applies \cite{Caticha00}, all that is needed is a
broader family of trial distributions.

Next we apply the extended ME method developed above to fluid Argon.

\section{An example: Argon}

One of the difficulties in testing theories about fluids against
experimental data is that it is not easy to see whether discrepancies are to
be blamed to a faulty approximation or to a wrong intermolecular potential.
This is why it is theories are normally tested against molecular dynamics
numerical simulations where there is control over the intermolecular
potential. In this section we compare ME results against simulation results 
\cite{Verlet68} for fluid of monatomic molecules interacting through a
Lennard-Jones potential (\ref{Lennard-Jones}). The parameters $\varepsilon $
and $\sigma $ (the depth of the well and the radius of the repulsive core, $%
u(\sigma )=0$) are chosen to model Argon: $\epsilon =10^{-2}\unit{eV}$ and $%
\sigma =3.405\unit{%
\text{\AA}%
}$.

\begin{figure}[tbp]
\includegraphics*[0in,0in][4in,3in]{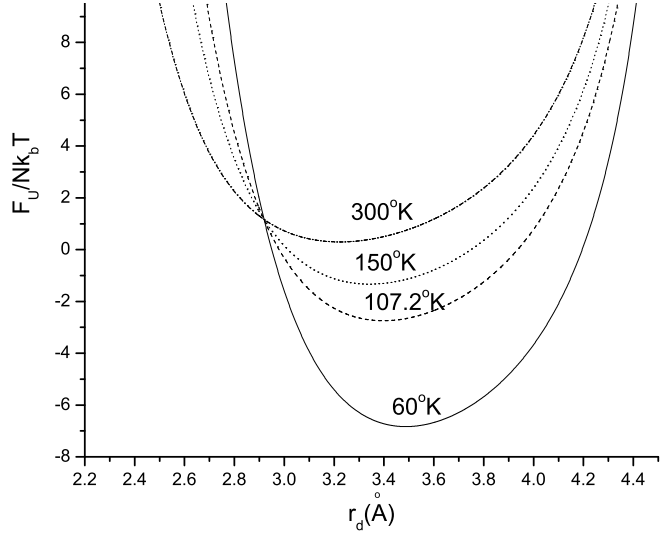}
\caption{ The free energy $F_{U}$ as a function of hard-sphere
diameter $r_{d}$ for Argon at a density of $\protect\rho \protect\sigma %
^{3}=0.65$ for different temperatures. The best $r_{d}$ is that which
minimizes $F_{U}$. }
\label{FudT}
\end{figure}

\begin{figure}[tbp]
\includegraphics*[0in,0in][4in,3in]{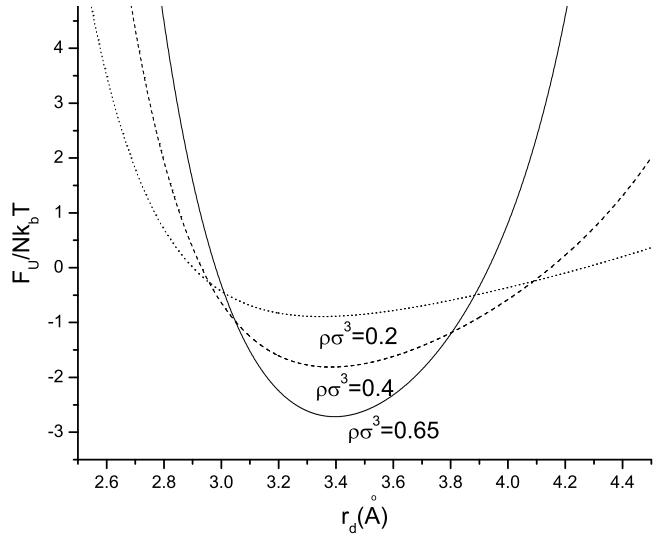}
\caption{ The free energy $F_{U}$ as a function
of hard-sphere diameter $r_{d}$ for Argon at $T=107.82\unit{K}$ for
different densities. The best $r_{d}$ is that which minimizes $F_{U}$.}
\label{Fudi}
\end{figure}

Figure \ref{FudT} shows the free energy $F_{U}/Nk_{B}T$ as a function of
hard-sphere diameter $r_{d}$ for Argon at a density of $\rho \sigma
^{3}=0.65 $ for different temperatures, and fig.\ref{Fudi} shows $%
F_{U}/Nk_{B}T$ as a function of $r_{d}$ for several densities at fixed $%
T=107.2\unit{K}$. Since the critical point for Argon is $T_{c}=150.69\unit{K}
$ and density $\rho \sigma ^{3}=0.33$ all these curves lie well within the
liquid phase. The increase of $F_{U}/Nk_{B}T$ for low values of $r_{d}$ is
due to short range repulsion described by $F_{hs}/Nk_{B}T$ while the
increase for large $r_{d}$ is due to the long range attraction described by $%
\langle U\rangle _{hs}/Nk_{B}T$.

The best $r_{d}$ is that which minimizes $F_{U}$ and depends both on
temperature and density. The best diameter decreases as the temperature
increases: atoms with higher energy can penetrate deeper into the repulsive
core. The dependence with density is less pronounced.

\begin{figure}[tbp]
\includegraphics*[0in,0in][4in,3in]{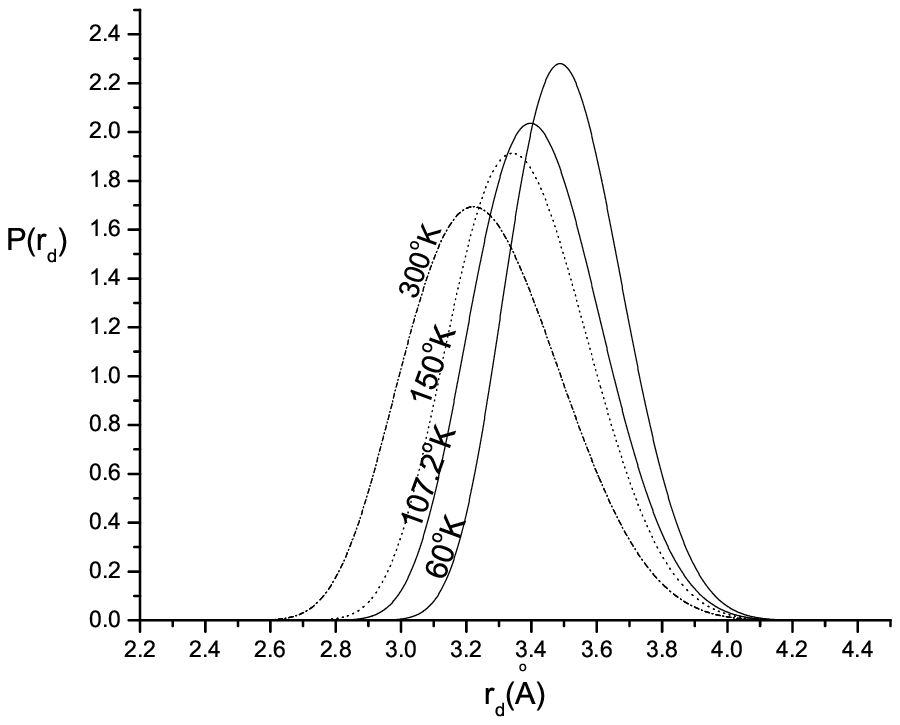}
\caption{ The distribution of hard-sphere
diameters $r_{d}$ for Argon for several temperatures at density $\protect%
\rho \protect\sigma ^{3}=0.65$. The size of fluid was arbitrarily chosen as $%
N_{\func{eff}}=13500$.}
\label{prddT}
\end{figure}

Next we study the distribution of diameters, eq.(\ref{Pd(rd)}). As discussed
earlier for large $N$ the distribution $P_{d}(r_{d})\sim \exp -\beta F_{U}$
is sharply peaked about the optimal diameter $r_{m}$. But we argued that the
effective number of particles that are relevant to the local behavior is
smaller $N_{\func{eff}}$. In Fig.\ref{prddT} we plot the distribution $%
P_{d}(r_{d})$ for different temperatures, for a fixed fluid density of $\rho
\sigma ^{3}=0.65$, and for an arbitrarily chosen fixed $N_{\func{eff}}=13500$%
. Notice again that the distribution shifts to lower diameters as the
temperature increases. Notice also that at lower temperatures the
distribution becomes narrower. This means that the effects of the softness
of the repulsive core are less important; that a hard-sphere approximation
is better at low temperatures \cite{BarkerHenderson76}.

\begin{figure}[tbp]
\includegraphics*[0in,0in][4in,3in]{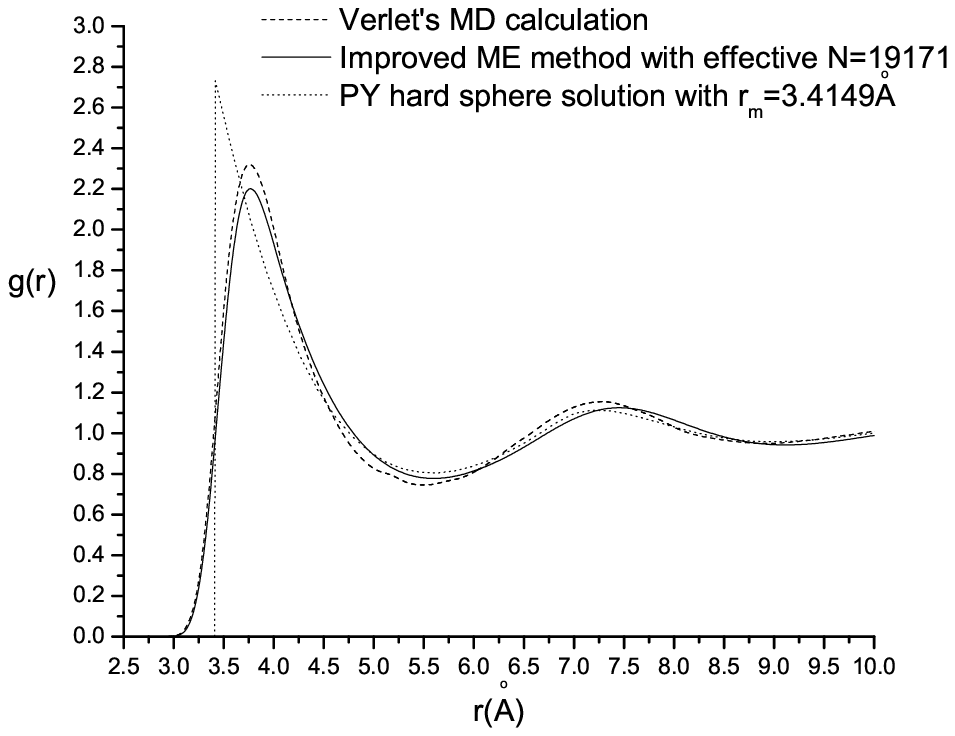}
\caption{The radial distribution function
for (a) the hard-sphere fluid with optimal diameter $r_{m}$; (b) Verlet's
molecular dynamics simulation; and (c) the improved ME analysis, for Argon
at density $\protect\rho \protect\sigma ^{3}=0.65$, temperature $T=107.82%
\unit{K}$, and effective particle number $N_{\func{eff}}=19171$.}
\label{gi065}
\end{figure}

Finally in fig.(\ref{gi065}) we show a comparison of the radial distribution
functions computed in three ways. The dotted line is $g_{hs}(r|r_{m})$ for
the hard-sphere fluid with optimal diameter $r_{m}$. This curve, calculated
from the numerical data tabulated in \cite{Throop65}, is also the result of
the variational method and coincides with the ME result for a
macroscopically large $N_{\func{eff}}=N$. The dashed line is Verlet's
molecular dynamics simulation The continuous line is the averaged $\bar{g}%
_{hs}(r)$ of the more extended ME analysis. The calculation is for a density 
$\rho \sigma ^{3}=0.65$, temperature $T=107.82\unit{K}$, and the value of
effective particle number $N_{\func{eff}}=19171$ was adjusted to achieve the
best fit. The agreement between the ME curve and Verlet's data is remarkably
good. The vast improvement over the simpler variational method calculation
is clear.

One might be tempted to dismiss this achievement as due to the adjustment of
the parameter $N_{\func{eff}}$ but this is not quite fair: there is a single
free parameter and the functional form of the whole curve $\bar{g}_{hs}(r)$
in eq.(\ref{g(r)-approx1}) is correctly reproduced. We should point out, for
example, that the Fisher-Rao measure $\gamma ^{1/2}\left( r_{d}\right) $
plays a crucial role. Omitting the factor $\gamma ^{1/2}\left( r_{d}\right) $
from the distribution $P_{d}(r_{d})$ would have led to a much less
satisfactory fit regardless of the choice of $N_{\func{eff}}$.

\section{Conclusion}

The main goal of this paper has been to show that the ME method can be used
to generate approximations in a way that generalizes the Bogoliuvob
variational principle. This addresses a range of applications that lie
beyond the scope of the traditional MaxEnt.

We showed that rather than approximating the real fluid by a fictitious one,
it is better to approximate the \textquotedblleft exact\textquotedblright\
probability distribution by a statistical mixture of distributions
corresponding to hard spheres of different diameters which allows a better
description of the soft core of the short-range repulsive potential.

The results achieved in this paper represent a step in the right direction
but they are by no means final. When we say that the averaged $\bar{P}%
_{hs}(q_{N})$ is the \textquotedblleft best\textquotedblright\ approximation
to $P(q_{N})$ we do not mean that one cannot do better. In fact, further
improvements are always achievable by choosing a broader family of trial
distributions. In the study of fluids we saw that these improvements are not
just possible, they are necessary. We argued that the next important
improvement of the ME method as a calculational tool for fluids should be in
the direction of developing a theory of local fluctuations. This would lead
to a systematic method for the determination of the effective number of
particles $N_{\func{eff}}$ that are locally relevant.

We conclude, therefore, that the ME approach to the theory of fluids is
already a significant improvement over the Bogoliuvob variational approach.
With only moderate further developments we can realistically expect it to
surpass even the best perturbative methods developed to date.

\section{Acknowledgments}

The authors acknowledge R. Scheicher and C.-W. Hong for their valuable
assistance and advice with the numerical calculations.

\end{document}